\documentclass[apjl]{emulateapj}

\newcommand{\etal}{et al.}

\newcommand{\cerenkov}{{\v C}erenkov}
\newcommand{\sgra}{Sgr A*}
\newcommand{\grs}{$\gamma$-rays}

\slugcomment{Accepted to ApJ Letters}

\shorttitle{The Galactic Center HESS Source and \sgra}
\shortauthors{Ballantyne \etal}

\usepackage{times}

\begin{document}

\title{A Possible Link Between the Galactic Center HESS Source and Sgr A*}


\author{D. R. Ballantyne\altaffilmark{1}, Fulvio
  Melia\altaffilmark{1,2,3}, Siming Liu,\altaffilmark{4} and Roland
  M. Crocker\altaffilmark{5}}

\altaffiltext{1}{Department of Physics, The University of Arizona, 1118 East 4th
  Street, Tucson, AZ 85721; drb, melia@physics.arizona.edu}
\altaffiltext{2}{Steward Observatory, The University of Arizona, 933
  N. Cherry Avenue, Tucson, AZ 85721}
\altaffiltext{3}{Sir Thomas Lyle Fellow and Miegunyah Fellow.}
\altaffiltext{4}{Theoretical Division, Los Alamos National Laboratory,
  P.O. Box 1663, Los Alamos, NM 87545; liusm@lanl.gov}
\altaffiltext{5}{School of Chemistry and Physics, The University of
  Adelaide, Adelaide, South Australia, 5005 Australia;
  roland.crocker@adelaide.edu.au}

\begin{abstract}
Recently, HESS and other air \cerenkov\ telescopes have detected a
source of TeV \grs\ coincident with the Galactic center. It is not yet
clear whether the \grs\ are produced via leptonic or hadronic
processes, so it is important to consider possible acceleration sites
for the charged particles which produce the \grs. One exciting
possibility for the origin of these particles is the central black hole, Sgr A*,
where the turbulent magnetic fields close to the event horizon can accelerate
protons to TeV energies. Using a realistic model of the density
distribution in a 6~pc$\times$6~pc$\times$6~pc cube at the Galactic center, we
here calculate the trajectories followed by these TeV protons as they
gyrate through the turbulent medium surrounding Sgr A*. Diffusing
out from the black hole, the protons produce TeV \grs\ via $\pi^0$ 
decay following a collision with a proton in the surrounding medium. After
following over 222,000 such trajectories, we find that the
circumnuclear ring around \sgra\ can reproduce the observed 0.1--100~TeV
HESS spectrum and flux if the protons are injected into this medium with
an effective power-law index $\approx 0.75$, significantly harder than the observed 
photon index of $2.25$. The total energy in the steady-state $1$--$40$ TeV 
proton population surrounding Sgr A* is inferred to be $\approx 5\times
10^{45}$ ergs. Only 31\% of the emitted 1--100~TeV protons encounter
the circumnuclear torus, leaving a large flux of protons that diffuse 
outward to contribute to the Galactic ridge emission observed
by HESS on scales of $\gtrsim 1^\circ$.
\end{abstract}


\section{Introduction}
\label{sect:intro}
The Galactic center is a complex environment, harboring what is
believed to be a supermassive black hole, Sagittarius A* (Sgr A*),
surrounded by a blend of new and old stellar populations, thermal and
nonthermal gas components, and a wide array of both compact and
diffuse high-energy sources (see \citealt{mf01} or \citealt{m07} for
recent reviews). In an exciting recent development, the Galactic
center has been identified as a source of TeV \grs\ by the air
\cerenkov\ telescopes HESS \citep{aha04,aha06b}, Whipple
\citep{kos04}, CANGAROO \citep{ts04} and MAGIC \citep{alb06}. While
many of the original observations initially measured different spectra
and luminosities for the Galactic center source \citep{aha04},
subsequent reprocessing and analysis has brought all the detections
into agreement \citep{kat05}. Here, we concentrate exclusively on the
HESS data as they provide the best constraints on the source
properties.

The signal from the Galactic center was detected by HESS in
observations conducted over two epochs (June-August 2003 and
March-September 2004; \citealt{aha04,aha06b}) with a combined
$\sim38\sigma$ excess above the background \citep{aha06b}. The
spectrum of this source is a power law with photon index
$2.25\pm0.10$, and the total flux above 1~TeV is $(1.87 \pm 0.30)
\times 10^{-8}$~m$^{-2}$~s$^{-1}$ \citep{aha06b}. As argued by
\citet{crocker05}, this HESS source does not appear to be coincident
with the EGRET Galactic center source 3EG~J1746-2851. There are two
reasons for this: first, the EGRET source excludes the Galactic center
at the $99.9\%$ confidence level \citep{hd02,pohl05}. In contrast, the HESS
source is coincident within $\sim 30\arcsec$ of \sgra, though with a
centroid displaced roughly $7\arcsec$ ($\sim 0.4$~pc) to the East of
the Galactic center \citep{aha06b}. Second, the EGRET spectrum
extrapolated into the HESS energy range over-predicts (by a factor
$\sim20$) the TeV $\gamma$-ray flux of the Galactic center
source. Although it is possible that both sources are associated with
the SNR Sagittarius A East, the HESS TeV source is probably associated
with the black hole itself, given its angular proximity to \sgra.

If the TeV \grs\ do originate from \sgra, then this presents somewhat
of a theoretical puzzle: given what we understand about \sgra\ and the
interactions with its nearby environment, this object should not be
directly producing a significant flux of TeV photons. Several possible
scenarios for producing these \grs\ have been proposed
\citep[e.g.,][]{ad04,an05a,an05b,ha06}, including that of
\citet{lmp06}. These last authors show that protons can be energized
to TeV energies by stochastic acceleration in a magnetically dominated
funnel close to the black hole ($\sim 20-30$~Schwarzschild radii). The
proton acceleration models of \citet{beck06} may also be active (or
even dominant) in \sgra. In either case, the relativistic protons will
then diffuse along magnetic fields from small scales to large radii
without experiencing adiabatic loss as is the case for ejections of
plasma blobs. As the protons diffuse out through the surrounding
medium, they may scatter with hydrogen nuclei in a shocked stellar
wind region, and with the circumnuclear disk surrounding the black
hole. Such pp scattering events produce $\pi^0$'s, which
subsequently decay into two photons. Thus, this model suggests that
protons random-walking their way into the gaseous few central parsecs
surrounding \sgra\ may ultimately be the source of the HESS-detected
TeV \grs\ \citep[e.g.,][]{an05b}.  It is this process that we aim to examine carefully in this
{\it Letter} by carrying out a highly-detailed analysis of the proton
transport and interaction, starting at the acceleration site near the
black hole and extending out as far as several parsecs. We assume a
distance to the Galactic center of $R_0=7.94$~kpc \citep{eis03}.

\section{Calculations}
\label{sect:calc}

\subsection{Outward Proton Diffusion from \sgra}
\label{sub:motion}
The motion of a relativistic proton moving through the ISM is governed
solely by the Lorentz force due to the interstellar magnetic field:
${d{\bf v}/dt}=\left({\bf v} \times {\bf \Omega} \right)/\gamma$,
where $\gamma$ is the Lorentz factor for a proton with velocity ${\bf
v}$ and ${\bf \Omega} \equiv e{\bf B}/mc$ is the gyrofrequency for a
particle with charge $e$ and mass $m$ under the influence of a magnetic
field ${\bf B}$. Thus, given a description for the magnetic field in the
computational volume, the trajectory of the proton can be followed
exactly. As there is no net electric field, the magnitude of the
velocity $v_0$ remains constant.

Proton trajectories are calculated in a 6~pc$\times$6~pc$\times$6~pc
cube consisting of $10^6$ equally spaced cells centered on the
Galactic center. The hydrogen density $n_{\mathrm{H}}$ within this
volume is taken from the results of simulations by
\citet{rock04}. These authors computed the density distribution in
this region caused by the interactions of stellar winds from the
surrounding young stars. In addition to the stellar wind gas, the
volume also contains a high-density `torus' with an inner radius of
1.2~pc and a thickness of 1~pc representing the observed
circumnuclear disk containing molecular gas.

The magnetic field is assumed to be generated everywhere in this region
with an intensity that is proportional to $n_{\mathrm{H}}$. The average 
density in the model stellar-wind gas (taken
to be any region where $n_{\mathrm{H}} < 3\times 10^3$~cm$^{-3}$) is $\left <
n_{\mathrm{H}}^{\mathrm{sw}} \right > = 121$~cm$^{-3}$, while it is
$\left < n_{\mathrm{H}}^{\mathrm{mt}} \right > =$ 233,222~cm$^{-3}$ within
the model molecular torus. Taking $kT=1.3$~keV as the average temperature of the
stellar-wind gas \citep{bag03,rock04}, 100~K for the temperature of
the molecular torus (\citealt{rock04} and references therein), and
assuming equipartition, the average field intensity $B_0$ is $\approx 3$~mG
in the stellar-wind region and $\approx 0.35$~mG within the torus. 
As the proton moves through different densities in the
computational domain, $B_0$ is scaled to the appropriate value,
indicated by the local gas density.

While the above procedure scales the magnetic field
intensity at any position in the grid, we also require a physical
description of the field direction at each point.  A useful
prescription for describing a turbulent magnetic field was developed
by \citet{gj94} to analyze the fundamental physics of ionic motion in
the ISM (see also \citealt{km00}). This quasi-analytic approach
assumes that the magnetic field fluctuations are static and follow a
Kolmogorov spectrum (though in principle any type of turbulence may
be modeled in this way). At every position in the grid,
the three-dimensional magnetic field is written as the sum
over a given number of transverse waves of random polarization, with a
wavevector oriented in a random direction. In contrast to
\citet{gj94}, we add together fluctuations from 200 wavelengths,
ranging from 0.1$v_0/\Omega_0$ to 10$v_0/\Omega_0$, where $\Omega_0$
is the proton gyrofrequency at $B_0$. In this fashion,
we may then generate all three components of the magnetic field as
functions of the position coordinates, for each proton released from
the origin with initial velocity ${\bf v}_0$ in a random direction.

\begin{figure*}
\includegraphics[width=0.57\textwidth]{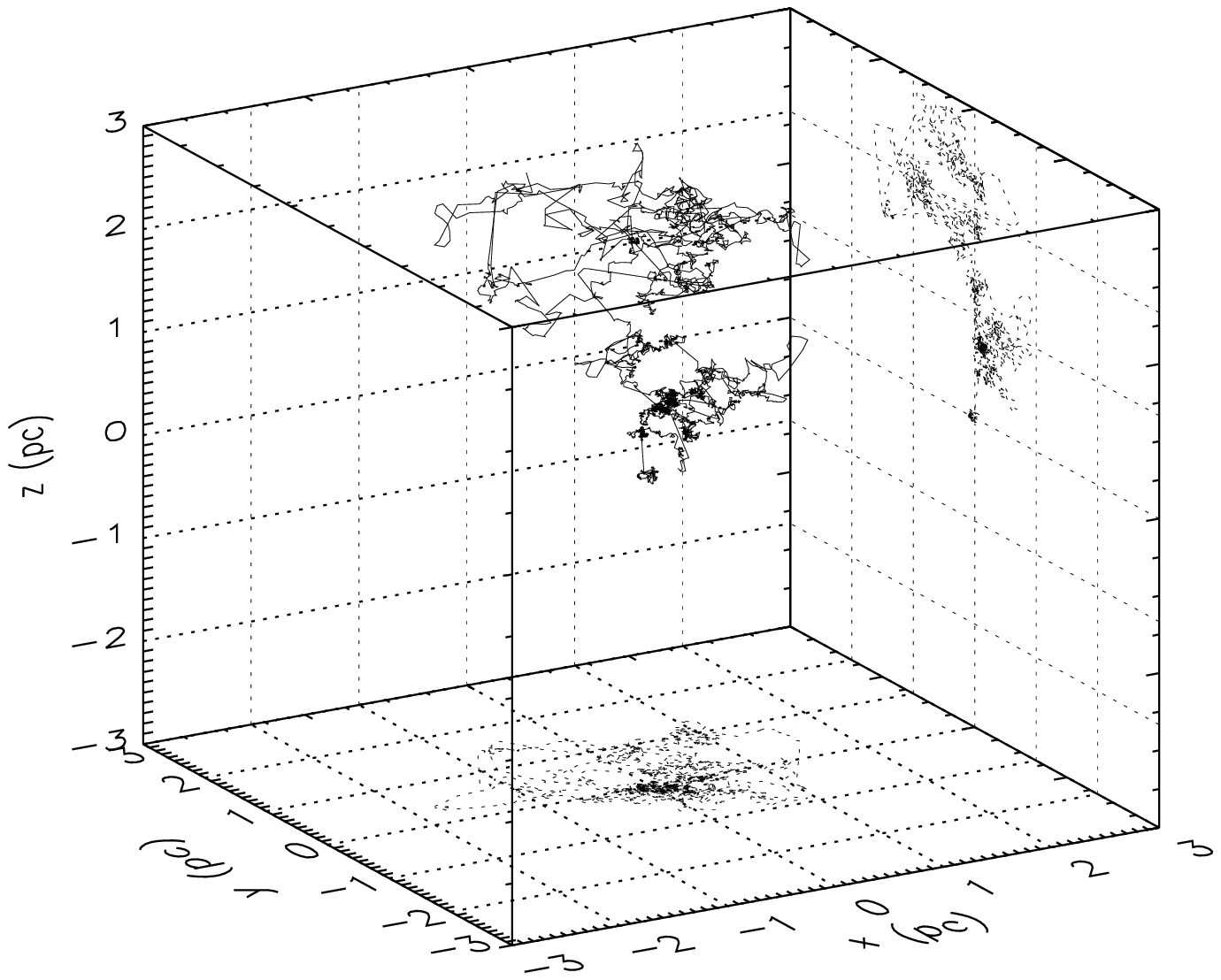}
\includegraphics[width=0.42\textwidth]{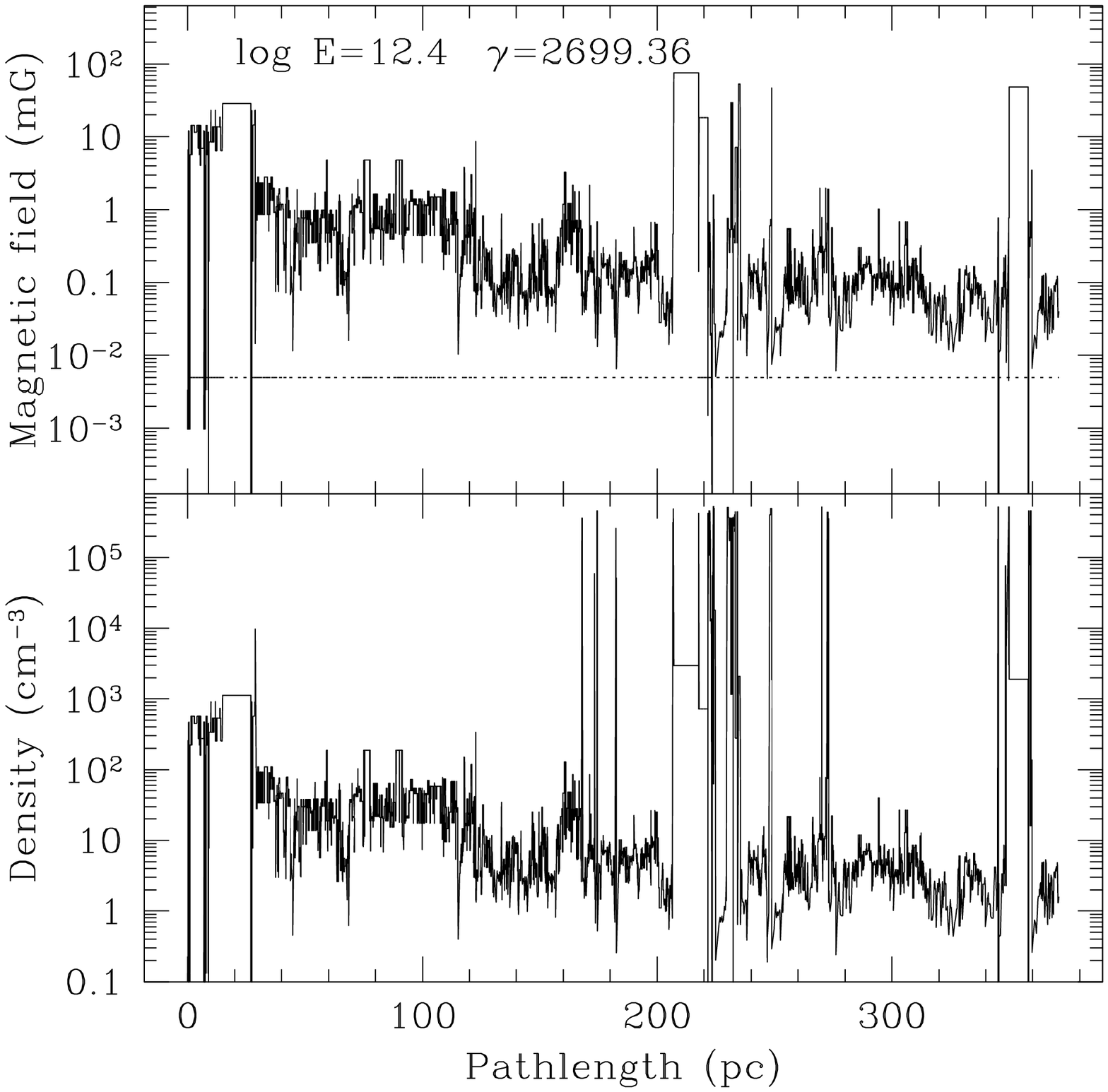}
\caption{(Left) Complete trajectory of a $\log (E/\mathrm{eV}) = 12.4$ proton from
  ejection at \sgra\ (the origin) until it escapes the computational volume at
  $z=3$~pc. The x-y plane is the sky plane. (Right) The magnetic
  field strength and gas density encountered by this proton on its
  trajectory. The particle traveled for 1,211~yrs and traversed
  371.3~pc on its journey outward from \sgra. It spent most of its
  time in the lower-density stellar wind gas but did have numerous
  encounters with the high density molecular torus. The dotted line in
  the upper-panel indicates the lowest magnetic field strength for
  which the Lorentz equation was solved. For the extremely low density
  zones which had magnetic field strengths below this limit the
  particle was moved in a straight line until it encountered higher
  density gas.}
\label{fig:example}
\end{figure*}

With the above ingredients, the trajectory of each proton can be
calculated exactly as it traverses the volume around \sgra. The
calculation is completed once the proton leaves the computational
domain. Since the observed-frame time step is proportional to
$\gamma^2$, computational resources limited the calculations to proton
energies between $1$ and $100$~TeV. A total of 222,617 proton
trajectories were calculated with energies uniformly distributed in
this range. Figure~\ref{fig:example} shows a typical trajectory
calculated in this fashion.

\subsection{The Pion Production Rate and Gamma-ray Emissivity}
As the proton random-walks its way toward the edge of the calculation
domain by scattering off the turbulent magnetic field, it may collide
with a low-energy proton in the ambient medium and produce pions via the
reaction $pp \rightarrow pp\pi^0\pi^\pm$, where a variable number of
pions are produced to conserve charge. The neutral pion will subsequently decay
into two photons. In principle, one or more of the two protons in each
scattering event may retain (or gain) sufficient energy to produce
additional pions in subsequent interactions, but we will ignore their
contribution to the overall pion production rate for this application. 
In addition, a charge exchange may occur (roughly $1/4$ of the time), 
in which a neutron is produced in the exit channel. A fitting formula 
for the cross-section of pp-scattering is given by \citet{eid04}, and 
is typically $\sigma_{pp} \sim 40$~mbarns for proton energies between 
1 and 10~TeV.

To compute the fraction of relativistic protons that undergo
pp-scattering, we first split the trajectories into $21$ energy bins
($\log (E/\mathrm{eV}) = 12, 12.1, \ldots, 13.9, 14$). Each proton
path is then followed and if the particle enters a volume element with
$n_{\mathrm{H}} > 3\times 10^3$~cm$^{-3}$ (these regions will dominate
the $\gamma$-ray emissivity) the incremental optical depth
$d\tau=\bar{n}_{\mathrm{H}} \sigma_{pp}(E) dl$ to pp-scattering
within this grid cell is calculated. Here, $dl$ is the distance the
proton traveled since the previous time step ($\approx 1/\Omega$) and
$\bar{n}_{\mathrm{H}}$ is the average density encountered by the
proton as it moved a distance $dl$. As long as the particle remains in
that cell, we continue to sum $d\tau$. If the proton scatters out of
the volume element and then subsequently returns, the new values of
$d\tau$ are added to the previous total. This procedure is followed
for each proton trajectory in each of the 21 energy bins. The fraction
of protons at energy $E$ that undergo a pp-scattering event in a
volume element with $n_{\mathrm{H}} > 3\times 10^3$~cm$^{-3}$ is
$f(E)=1-e^{-(\int d\tau)/N}$, where $N$ is the number of calculated
trajectories at energy bin $E$.

A power-law spectrum of proton energies in the relativistic regime is
a natural consequence of many acceleration processes
\citep[e.g.,][]{lmp06}, but the exact value of the spectral index
depends on the details of how the protons escape. As we shall see
below, the rate of diffusion and subsequent scattering of the protons
is highly energy dependent. Therefore, we treat the
``injected" spectral index as a variable to be fixed by the fitting
procedure. By ``injected" we mean the distribution of protons leaving
Sgr A*'s region of influence and entering the wind-shocked region
surrounding it.

The proton spectrum is thus written as $dn/dE_{\mathrm{inj}} = K
(E/E_{\mathrm{min}})^{-\alpha}$, where
$K\propto \mathcal{E}$, the total energy in protons in ergs between
$E_{\mathrm{min}}=1$~TeV and $40$~TeV. The
spectrum of scattered protons in a grid cell is thus $dn/dE= \left(
dn/dE_{\mathrm{inj}} \right ) f(E)$, where $f(E)$, defined above, is
the fraction of protons with energy $E$ which collide with an ambient
proton in that volume element. However, $f(E)$ is only known down to
$\log E=12$, so $dn/dE$ is extrapolated to $0.1$~TeV using the chosen
power-law index $\alpha$ and the value of $f(E)$ at $\log E=12$. As
\grs\ at a given energy are preferentially produced by protons with
$\sim 10\times$ greater energy, this extrapolation will not greatly
affect the predictions between 0.1 and 1~TeV. The fraction of protons
that interact with the circumnuclear torus becomes very small at
energies greater than 100~TeV (see \S~\ref{sect:results}), so it is
not necessary to extrapolate $dn/dE$ beyond 100~TeV. Now armed with
the $dn/dE$ profile within each cell, we calculate the $\pi^0$ and
$\gamma$-ray emissivity for different values of $\mathcal{E}$ and
$\alpha$ using the formalism described by \citet{crocker05}.  The
emissivities are then converted into fluxes before the values are 
summed to produce the predicted $\gamma$-ray spectrum from
0.1--100~TeV.

\section{Results}
\label{sect:results}
\begin{figure}
\includegraphics[angle=-90,width=0.5\textwidth]{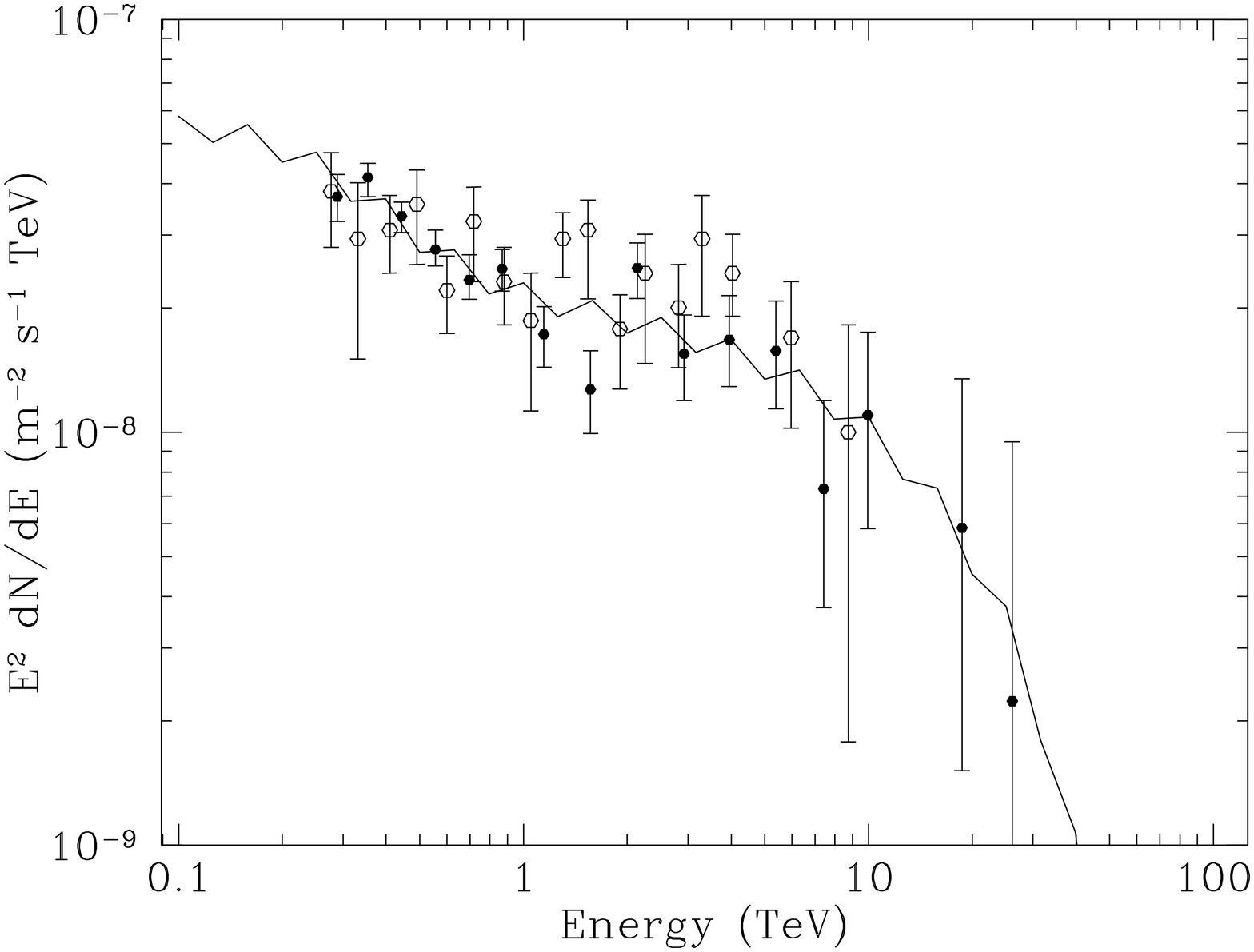}
\includegraphics*[width=0.505\textwidth,viewport=70 10 500 350]{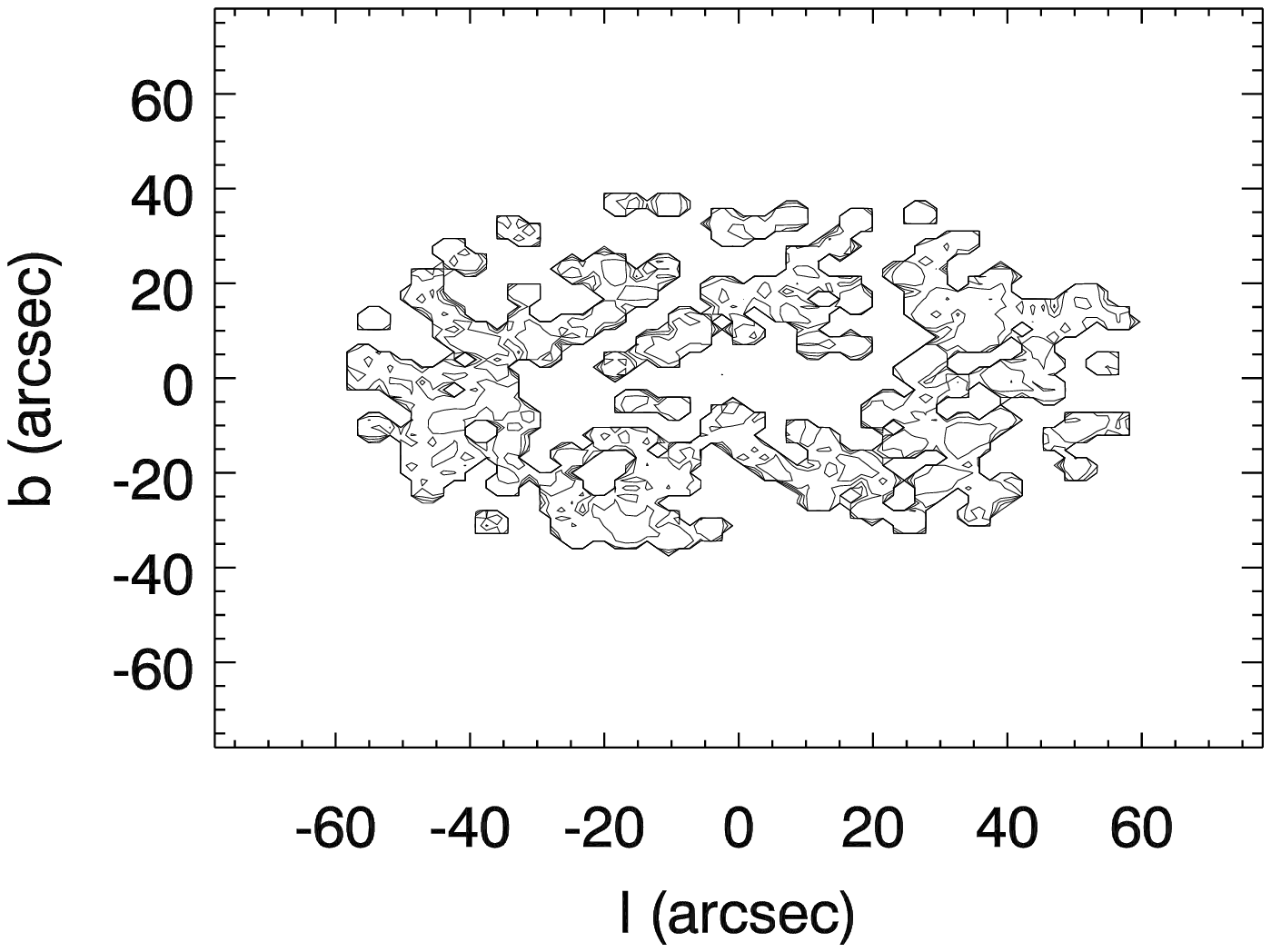}
\caption{(Top) The solid line is the predicted spectrum from the
  Galactic center region due to pp-scattering. The total energy
  required in protons between 1 and 40~TeV is $5\times 10^{45}$~ergs,
  and the injected spectrum is $dn/dE_{\mathrm{inj}} \propto
  E^{-0.75}$. The open data points are the 2003 observed values from
  HESS taken from \citet{aha04}, while the solid points are from the
  2004 HESS observations \citep{aha06b}. (Bottom) Contours of 1--10~TeV
  flux predicted from the same model that produced the spectrum in the
  left-hand panel. The contours are at $\log f =
  -15,-14,-13,-12,-11,-10,$ and $-9$ [s$^{-1}$~m$^{-2}$]. The
  oblateness evident in this image is due to the tilt angle of the
  molecular torus. \sgra\ is at the origin.}
\label{fig:spect}
\end{figure}

We compared the model spectra to the observed 2003 and 2004 HESS
data between 0.3 and 30~TeV (taken from \citealt{aha04,aha06b}) 
using $\chi^2$ fitting. The lowest $\chi^2$ was found for $\mathcal{E} 
= 5\times 10^{45}$~ergs and $\alpha=0.75$ [$\chi^2/$d.o.f$=41.7/31$ 
($29.1/31$) using the smaller (larger) of the observed error bars, and 
d.o.f. = degrees of freedom]. This value for the spectral index is
very robust with $\Delta \chi^2 \approx +5$ for $\alpha=0.7$ or
$0.8$. The predicted 0.1--100~TeV spectrum and 
1--10~TeV image of the Galactic center with these parameters are shown 
in Fig.~\ref{fig:spect}. The expected 1--10~TeV luminosity from this 
spectrum is $4.7\times 10^{34}$~erg~s$^{-1}$. The predicted 1--100~TeV 
flux is $1.83\times 10^{-8}$~m$^{-2}$~s$^{-1}$, in good agreement with 
the observed value of $(1.87 \pm 0.30) \times 10^{-8}$~m$^{-2}$~s$^{-1}$ 
\citep{aha06b}.

The fact that the best fit corresponds to $\alpha=0.75$ indicates that
the spectrum of injected particles must be significantly harder than
the observed TeV $\gamma$-ray spectrum (for which the spectral index
is $2.25\pm0.10$). This is simply because the higher energy protons
diffuse outward from the Galactic center faster than the low energy
ones and as a result traverse, on average, a much smaller path
length. These particles are then less likely to encounter the
molecular torus and undergo a pp-scattering event that will produce a
TeV $\gamma$-ray. A harder intrinsic spectrum is therefore required to
compensate for this deficit and account for the observed spectral
shape. The derived value of $\alpha=0.75$ is much harder than the
``standard" power law with index $\sim 2-2.5$ that one might expect
from simple first-order Fermi scattering. Perhaps this may be taken as
an indication that Sgr A* is not the source of energetic hadrons
producing the $\gamma$-ray glow in its vicinity. Alternatively, the
intrinsic $\alpha \sim 2$ proton spectrum may have been altered by
interactions much closer to the black hole. Another possibility is
that the required hard proton spectrum might be an indication that
stochastic acceleration continues as the particles diffuse outwards
from the initial injection region. Future work will investigate the
origin and viability of the hard proton spectrum.

We find that only about 30\% of the computed trajectories encounter
the circumnuclear torus; the rest travel unimpeded to larger
distances.  However, this fraction is highly energy dependent:
$\sim73$\% of the emitted protons encounter the torus at $\log
(E/\mathrm{eV})=$12--12.4, but this drops to 47\% at $\log
(E/\mathrm{eV})=13$, and finally to 5\% at $\log
(E/\mathrm{eV})=14$. As a result, the predicted $\gamma$-ray spectrum
exhibits a roll-over at energies $\gtrsim 20$~TeV
(Fig.~\ref{fig:spect}). This is a robust prediction of models where
\sgra\ is the source of the relativistic protons and can be tested
with future HESS observations. 

The energy-dependent proton interaction rate also suggests that a
significant flux of TeV protons escapes from the inner few pc
surrounding \sgra\ to interact with more distant molecular gas.
This would be consistent with the observed emission from the Galactic
center ridge. \citet{aha06a} report 
that the inner 200~pc of the Galaxy glows in TeV \grs, with the dominant
emission correlated with the distribution of dense molecular gas.
In addition, the spectrum of the Galactic ridge emission is observed to 
be close to that of the Galactic center source. The best-fitting model 
for the Galactic center HESS source predicts an energy of $\mathcal{E}= 
5\times 10^{45}$~ergs in protons between 1 and 40~TeV. \citet{aha06a} 
estimate that a hadron distribution with an energy $\sim 10^{49}$~ergs 
(from 4--40~TeV) is required to produce the observed Galactic
ridge emission. The ridge region is much larger than the volume within
which the Galactic center TeV source is produced, so the escaping
protons would build up the much larger energy content over time.
The diffusion time to reach such large distances is $> 10^4$~years 
\citep{aha06a}, so the particle accelerator would have to be at
least that old. 

Our simulations show that an injected proton index $\alpha\approx 0.75$ 
is required to produce the observed $\gamma$-ray index $\alpha_{\mathrm{obs}} 
\approx 2.25$, a steepening of $3/2$. Homogeneous diffusion with a Kolmogorov 
or Bohm turbulence spectrum would produce a steepening of $1/3$ and $1$, 
respectively. Our treatment in this paper has followed the diffusion of protons 
taking into account the variable 3-D gas density and magnetic field 
surrounding \sgra. On the other hand, diffusion calculations assume 
a homogeneous medium, which produces misleading results when applied 
to regions as complex as that near \sgra. As such, our finding that
a steepening of 3/2 is required, instead of $\sim 1$, may be traceable 
to our more accurate treatment of inhomogeneity. 

\section{Conclusions}
\label{sect:concl}
The HESS observations suggest that there may be a common cause for 
both the Galactic-center point source and the diffuse emission along the
ridge. TeV protons accelerated near \sgra\ can account for the observed 
flux and spectrum of the central source through scattering events with
ambient protons within the circumnuclear ring. The proton spectrum
injected into the gas surrounding \sgra\ must have a hard power-law
index of $0.75$, implying a non-standard acceleration process or
interactions close to the acceleration region. Our simulations have
resulted in a predicted 1--10~TeV image of the source for comparison 
with future high-resolution observations. We have found, in particular, 
that the predicted TeV image mirrors closely the structure of the torus, 
including an oblateness due to projection effects in the plane of the 
sky. The total energy required in protons between 1 and 40~TeV is $5\times
10^{45}$~ergs and a significant fraction of the protons do escape the
neighborhood of \sgra\ without undergoing a pp collision; these
presumably diffuse to much larger distances where they can interact
with other molecular material, possibly accounting for the observed 
Galactic ridge emission. Although future calculations are required 
to confirm this suggestion, and other origins for the observed HESS 
source are possible \citep[e.g.,][]{wlg06,ha06}, this work supports 
the view that \sgra\ may be an important site for particle acceleration 
at the center of the Galaxy.

\acknowledgments

We thank the anonymous referee for very useful comments.
DRB is supported by the University of Arizona Theoretical Astrophysics
Program Prize Postdoctoral Fellowship. This work was funded in part at
the University of Arizona by NSF grant AST-0402502, and has made use
of NASA's Astrophysics Data System Abstract Service.  FM is
grateful to the University of Melbourne for its support (through a Sir
Thomas Lyle Fellowship and a Miegunyah Fellowship).

{}

\end{document}